\documentclass[prl,twocolumn,superscriptaddress]{revtex4-2}

\usepackage[T1]{fontenc}
\usepackage{amsmath,amssymb,bm}
\usepackage{graphicx}
\usepackage{dcolumn}
\usepackage[dvipsnames]{xcolor}
\usepackage[
  colorlinks=true,
  citecolor=Green,
  linkcolor=Green,
  urlcolor=Green
]{hyperref}
\begin{document}

\title{Circuit Complexity of Scalar and Tensor Perturbations in Luminal Einstein--Gauss--Bonnet Inflation}
\author{Tao Li}
\email{2022700450@stu.jsu.edu.cn}
\author{Hai-Bing Fu}
\email{fuhb@gzmu.edu.cn}
\affiliation{Department of Physics, Guizhou Minzu University, Guiyang 550025, China}
\author{Lei-Hua Liu}
\email{liuleihua8899@hotmail.com}
\affiliation{Department of Physics, College of Physics, Mechanical and Electrical Engineering, Jishou University, Jishou 416000, China}

\begin{abstract}
We investigate the circuit complexity of primordial scalar and tensor perturbations in the luminal branch of Einstein–Gauss–Bonnet (EGB) inflation. We construct the Hamiltonians from the quadratic actions and derive the evolution equations for the squeezing parameters. Although the tensor propagation speed is fixed to unity, the Gauss–Bonnet (GB) coupling remains dynamical through the scalar and tensor pump fields. We then calculate the Nielsen circuit complexity of the corresponding two-mode squeezed states. The GB correction produces a clear separation between the squeezing amplitudes, while the squeezing angles remain nearly phase locked. Therefore, the circuit complexity is mainly controlled by the evolution of the squeezing amplitudes. For the benchmark parameters considered here, a positive coupling increases the preparation cost of the scalar Gaussian state but slightly decreases that of the tensor Gaussian state. The scalar sector is therefore more sensitive to the GB coupling, whereas the luminal tensor sector provides a cleaner probe of the pump-field mechanism. Their opposite responses provide a differential state-space diagnostic of the same EGB background. 
\end{abstract}

\maketitle

\section{Introduction}
\label{sec:introduction}

Circuit complexity quantifies the minimum cost required to prepare a target quantum state from a reference state. Its geometric formulation converts this optimization problem into the search for a shortest path on a unitary manifold~\cite{Nielsen:2005mkt,Dowling:2006tnk}. It has been extended to Gaussian quantum field theories through wave-function and covariance-matrix methods~\cite{Jefferson:2017sdb,Chapman:2017rqy,Chapman:2018hou,Hackl:2018ptj}. These approaches have clarified how the result depends on the reference state, gate set, cost function, and penalty factors~\cite{Bhattacharyya:2018bbv,Alves:2018qfv,Camargo:2018eof,Ali:2018fcz}. Cosmological perturbations provide a natural setting because expansion converts the initial vacuum into a two-mode squeezed state~\cite{Grishchuk:1990bj,Albrecht:1992kf}. Circuit complexity can then measure the cost of preparing the evolved perturbation state and distinguish histories that lead to similar late-time spectra~\cite{Bhattacharyya:2020art,Lehners:2020pem}. The full Gaussian state contains information that is not fixed by the power spectrum alone. Previous inflationary analyses show that the dominant state-space separation is governed by the growth of squeezing amplitude $r_k$, whereas squeezing angle $\varphi_k$ supplies the phase information required by phase-sensitive complexity measures
\cite{Liu:2021nzx,Li:2023ekd,Li:2024kfm,Li:2026cgb,Li:2026jxx,Li:2021kfq}. 

We use this physical interpretation to determine how the EGB background changes the preparation cost of scalar and tensor Gaussian states. During inflation, the squeezing amplitude controls the occupation of correlated mode pairs, while the squeezing angle records their phase-space orientation. Large superhorizon squeezing can make equal-time correlators resemble a classical stochastic distribution, although the underlying opposite-momentum modes retain quantum correlations~\cite{Martin:2015qta}. A state-preparation measure is therefore useful because it responds to both the amplification and the phase structure of the perturbations. Related operator-based constructions likewise show that circuit complexity is sensitive to squeezing and to the time-dependent rotation of the state~\cite{Haque:2021hyw}. This motivates treating complexity as a quantum-information diagnostic that complements, rather than replaces, conventional spectra and correlation functions.

The EGB inflation couples the inflation to the curvature invariant $\mathcal G=R^2-4R_{\mu\nu}R^{\mu\nu}+R_{\mu\nu\rho\sigma}R^{\mu\nu\rho\sigma}$. Because a constant GB coefficient is topological in four dimensions, a scalar-dependent coupling is required to modify the background and perturbations. Such couplings arise in low-energy effective descriptions of gravity and support slow-roll, constant-roll, and graceful-exit inflationary solutions~\cite{Kawai:1999pw,Satoh:2008ck,Guo:2009uk,Akbarieh:2021vhv,Guo:2010jr,Kanti:2015pda,Yi:2018gse}. Their effects on the scalar and tensor spectra, spectral indices, and tensor-to-scalar ratio have been studied analytically and numerically in a broad class of potentials and coupling functions~\cite{Koh:2014bka,Yi:2018gse,Mudrunka:2023wxy}. In particular, the coupling modifies the usual inflationary consistency relation and may generate tensor spectra that are not available in minimally coupled slow-roll models~\cite{Satoh:2010ep}. The same interaction is contained in the general second-order scalar--tensor framework, where the quadratic scalar and tensor actions provide the no-ghost and gradient-stability conditions~\cite{Kobayashi:2011nu}.

At quadratic order, the coupling modifies the scalar kinetic coefficient and sound speed as well as the tensor kinetic coefficient and tensor speed~\cite{Hwang:2005hb,DeFelice:2011uc,Jiang:2013gza}. A viable perturbative background must keep both kinetic coefficients and both squared propagation speeds positive throughout the evolution. Requiring luminal tensor propagation imposes an additional relation between the coupling function and the inflationary background, while still allowing non-trivial scalar and tensor pump fields~\cite{Odintsov:2020sqy}. We therefore focus on the luminal branch $c_T^2=1$. This condition removes the direct tensor-speed correction, but the tensor kinetic coefficient $q_T$ and pump field $z_T'/z_T$ remain time dependent. The scalar sector is modified independently through $Q_s$, $c_s$, and $z_s'/z_s$. Consequently, the two sectors provide distinct probes of the same higher-curvature background even when their tensor propagation speed is fixed to unity. Scalar and tensor modes evolve in the same GB background but follow different squeezing trajectories. The power spectra describe the amplitudes of the perturbations, while circuit complexity measures the distance from the initial vacuum to the complete Gaussian target state. The latter retains information from both the squeezing amplitude and the squeezing angle. Comparing the two sectors therefore tests whether a luminal modified-gravity background produces different quantum-state preparation costs. We derive the two-mode Hamiltonians for the scalar and tensor perturbations, solve the corresponding squeezing equations, and substitute the results into a common geometric \(F_2\) wave-function cost. This comparison reveals the different roles of the two pump fields and provides a state-space diagnostic that complements conventional perturbation observables.

The paper is organized as follows. Sec.~\ref{sec:GB_compact_theory} presents the EGB quadratic actions, two-mode Hamiltonians, and scalar and tensor squeezing dynamics. Sec.~\ref{sec:circuit_complexity} introduces the wave-function circuit complexity and applies it to the numerical squeezing solutions. Sec.~\ref{sec:conclusion} summarizes the results and discusses their scope.

\section{Scalar and tensor squeezing in EGB inflation}
\label{sec:GB_compact_theory}

We consider a canonical inflation coupled to the GB invariant
\begin{equation}
\begin{split}
     &S=\int d^4x\sqrt{-g}\left[ \frac{M_{\rm Pl}^2}{2}R-\frac12(\partial\phi)^2
 -V(\phi)-\xi(\phi)\mathcal G \right],\\&~~~~~~~~~~~~~~~\mathcal G=R^2-4R_{\mu\nu}R^{\mu\nu}+R_{\mu\nu\rho\sigma}R^{\mu\nu\rho\sigma}.
\end{split}
 \label{eq:compact_GB_action}
\end{equation}
The background equations and the perturbation constraints have been derived in Refs.~\cite{Guo:2009uk,Guo:2010jr,Jiang:2013gza}.  We use their final perturbation coefficients.  Our coupling convention is $-\xi\mathcal G$. For the convention $-\widetilde\xi\mathcal G/2$, one must set $\widetilde\xi=2\xi$.

\subsection{Quadratic actions and two-mode Hamiltonians}
\label{subsec:GB_compact_Hamiltonians}

The scalar quadratic action is
\begin{equation}
 S_s^{(2)}=\frac12\int dt\,d^3x\,a^3Q_s
 \left[\dot{\mathcal R}^{2}
 -\frac{c_s^2}{a^2}(\partial_i\mathcal R)^2\right].
 \label{eq:compact_scalar_action}
\end{equation}
This result follows from the ADM action after the lapse and shift constraints are eliminated~\cite{Hwang:2005hb,DeFelice:2011uc,Jiang:2013gza}.  Define $\Delta=8H\dot\xi/(M_{\rm Pl}^2-8H\dot\xi)$.  In our convention, the two coefficients are
\begin{align}
 &~~~~~~~~~~~Q_s=\frac{\dot\phi^2+12\Delta H^3\dot\xi} {H^2(1-\Delta/2)^2},
 \label{eq:compact_Qs}
 \\&
 c_s^2=1+ \frac{16\Delta H\dot\xi\dot H +4\Delta^2H^2(\ddot\xi-H\dot\xi)}
 {\dot\phi^2+12\Delta H^3\dot\xi}.
 \label{eq:compact_cs}
\end{align}
The quantities $Q_s$ and $c_s$ are fixed by the background solution $H(t)$ and $\phi(t)$, with $\dot{\xi}=\xi_{,\phi}\dot{\phi}$ and $\ddot{\xi}=\xi_{,\phi\phi}\dot{\phi}^{2}+\xi_{,\phi}\ddot{\phi}$. Thus, $c_s$ is evaluated from the background rather than prescribed independently.  For the tensor perturbation, we use $h_{ij}=a^2(e^\gamma)_{ij}$ with $\partial_i\gamma_{ij}=0$ and $\gamma_{ii}=0$.  The known quadratic action is
\begin{equation}
 S_T^{(2)}=\frac18\int dt\,d^3x\,a^3q_T
 \left[\dot\gamma_{ij}^{2}
 -\frac{c_T^2}{a^2}(\partial_\ell\gamma_{ij})^2\right],
 \label{eq:compact_tensor_action}
\end{equation}
where
\begin{equation}
 q_T=M_{\rm Pl}^2-8H\dot\xi,
 \qquad
 c_T^2=\frac{M_{\rm Pl}^2-8\ddot\xi}{M_{\rm Pl}^2-8H\dot\xi}.
 \label{eq:compact_qT_cT}
\end{equation}
Both coefficients are fixed by the same inflationary background. To separate the tensor pump-field correction from a modified tensor propagation speed, we restrict the numerical analysis to the luminal tensor-speed branch
\begin{equation}
 \ddot\xi=H\dot\xi,
 \qquad c_T^2=1.
 \label{eq:compact_luminal_branch}
\end{equation}
This is a theoretical restriction on the inflationary background rather than an inference from a late-time observation. The GB correction remains nontrivial because $q_T$ can vary with time and modify the tensor pump field. These coefficients are standard results of GB perturbation theory~\cite{Guo:2009uk,Guo:2010jr,Jiang:2013gza}. The conditions $Q_s>0$, $c_s^2>0$, $q_T>0$, and $c_T^2>0$ exclude ghost and short-scale gradient instabilities. For later use, define
\begin{equation}
 \delta_{\rm GB}=\frac{8H\dot\xi}{M_{\rm Pl}^2},
 \qquad
 \Delta=\frac{\delta_{\rm GB}}{1-\delta_{\rm GB}},
 \qquad
 \epsilon=-\frac{\dot H}{H^2}.
 \label{eq:compact_deltaGB}
\end{equation}
In addition, a star denotes evaluation at the horizon crossing of the reference mode. Therefore, \(\delta_{{\rm GB},*}\) represents the GB coupling strength when the reference mode crosses the horizon. Using the background equations on the luminal branch, the perturbation coefficients may be written as
\begin{align}
 &Q_s=M_{\rm Pl}^2\frac{2\epsilon(1-\delta_{\rm GB})+
 3\delta_{\rm GB}^2/[2(1-\delta_{\rm GB})]}
 {(1-\Delta/2)^2},
 \label{eq:compact_Qs_delta}
 \\
& c_s^2=1-\frac{2\epsilon\delta_{\rm GB}^2/(1-\delta_{\rm GB})}
 {2\epsilon(1-\delta_{\rm GB})+3\delta_{\rm GB}^2/[2(1-\delta_{\rm GB})]},
 \label{eq:compact_cs_delta}
 \\
& q_T=M_{\rm Pl}^2(1-\delta_{\rm GB}),
 \qquad c_T^2=1.
 \label{eq:compact_qT_delta}
\end{align}
We now use conformal time $d\eta=dt/a$.  For the scalar mode, define $v_s=z_s\mathcal R$, $z_s=a\sqrt{Q_s}$, and $h_s=z_s'/z_s$.  For each tensor polarization, define $v_T=z_T\gamma_\lambda$, $z_T=a\sqrt{q_T/2}$, and $h_T=z_T'/z_T$.  Both sectors can then be written as
\begin{equation}
 S_A^{(2)}=\frac12\int d\eta\,d^3x
 \left[(v_A'-h_Av_A)^2-c_A^2(\partial_iv_A)^2\right].
 \label{eq:compact_unified_action}
\end{equation}
The index $A=s,T$ labels the two sectors and does not introduce an additional propagation speed. Specifically, $(v_A,z_A,h_A,c_A)$ denotes $(v_s,z_s,z_s'/z_s,c_s)$ for scalar perturbations and $(v_T,z_T,z_T'/z_T,c_T)$ for tensor perturbations. Thus, the numerical evolution uses the distinct pairs $(c_s,z_s'/z_s)$ and $(c_T,z_T'/z_T)$. The conjugate momentum is $\pi_A=v_A'-h_Av_A$, and the Hamiltonian is
\begin{equation}
 H_A=\frac12\int d^3x\left[\pi_A^2+c_A^2(\partial_iv_A)^2 +h_A(\pi_Av_A+v_A\pi_A)
 \right].
 \label{eq:compact_field_Hamiltonian}
\end{equation}
We use a fixed reference frequency $k$.  This is the convention adopted for the squeezed state construction in Ref.~\cite{Li:2023ekd}.  The modes $ k$ and $- k$ form one two-mode sector.  Its Hamiltonian is
\begin{equation}
    \begin{split}
         \hat H_{A,\bm k}&=\Omega_{A,k} \left( \hat c_{A, k}^{\dagger}\hat c_{A, k} +\hat c_{A, k}^{\dagger}\hat c_{A,- k}+1 \right)\\&
+\Lambda_{A,k} \hat c_{A,k}^{\dagger}\hat c_{A,-k}^{\dagger} +\Lambda_{A,k}^{*} \hat c_{A,k}\hat c_{A,- k},
 \label{eq:compact_two_mode_Hamiltonian}
    \end{split}
\end{equation}
\begin{equation}
 \Omega_{A,k}=\frac{k}{2}(1+c_A^2),
 \quad
 \Lambda_{A,k}=\frac{k}{2}(c_A^2-1)+ih_A.
 \label{eq:compact_Omega_Lambda}
\end{equation}
The coefficient \(\Lambda_{A,k}\) controls the creation and annihilation of correlated mode pairs. Its real part, \(k(c_A^2-1)/2\), arises from a non-unity propagation speed, while its imaginary part, \(h_A=z_A'/z_A\), describes the parametric amplification generated by the evolving background. The propagation speed changes the real part of $\Lambda_{A,k}$.  The pump field changes its imaginary part.

The scalar coefficients are obtained by setting \(c_A=c_s\) and \(h_A=z_s'/z_s\), while the tensor coefficients are obtained by setting \(c_A=c_T\) and \(h_A=z_T'/z_T\). Even when \(c_T^2=1\), \(q_T\) can remain time dependent, and \(h_T\) can differ from \(a'/a\). The scalar pump field \(z_s=a\sqrt{Q_s}\) is the GB generalization of \(z\). In the general-relativistic (GR) limit, \(Q_s\to2M_{\rm Pl}^2\epsilon\) and \(z_s\to aM_{\rm Pl}\sqrt{2\epsilon}=z\). By contrast, \(z_T=a\sqrt{q_T/2}\) is controlled by the tensor kinetic coefficient. Therefore, \(z_s'/z_s\) and \(z_T'/z_T\) generally produce different squeezing dynamics, even on the luminal branch \(c_T^2=1\). In particular, using \(d\delta_{\rm GB}/dN=(1-\epsilon)\delta_{\rm GB}\) on this branch gives
\begin{align}
 \frac{z_s'}{z_s}&=aH\left(1+\frac12\frac{d\ln Q_s}{dN}\right),
 \label{eq:compact_hs_delta}
 \\
 \frac{z_T'}{z_T}&=aH\left[1-\frac{(1-\epsilon)\delta_{\rm GB}}
 {2(1-\delta_{\rm GB})}\right].
 \label{eq:compact_hT_delta}
\end{align}
The tensor squeezing can therefore differ from the GR result through the pump field, without invoking a nonluminal tensor speed.

\subsection{Two-mode squeezed states and squeezing parameters}
\label{subsec:GB_compact_squeezing}

The pair-creation terms in Eq.~\eqref{eq:compact_two_mode_Hamiltonian} generate a two-mode squeezed state.  For each sector, the squeeze operator is
\begin{equation}
\begin{split}
     \hat S_{A,\bm k}(r_{A,k},\varphi_{A,k})
 &=\exp\Bigg\{r_{A,k}\bigg[
 e^{-2i\varphi_{A,k}}
 \hat c_{A,\bm k}\hat c_{A,-\bm k}
 \\&-e^{2i\varphi_{A,k}}
 \hat c_{A,-\bm k}^{\dagger}
 \hat c_{A,\bm k}^{\dagger}
 \bigg]\Bigg\}.
 \label{eq:compact_squeeze_operator}
\end{split}
\end{equation}
Acting on the two-mode vacuum state gives
\begin{equation}
\begin{split}
     \lvert\psi_A(\eta)\rangle_{\rm sq}
& =\frac{1}{\cosh r_{A,k}}\\&
 \times\sum_{n=0}^{\infty}
 \left[-e^{2i\varphi_{A,k}}\tanh r_{A,k}\right]^n
 \lvert n;n\rangle_{A,k,-k}.
 \label{eq:compact_squeezed_state}
\end{split}
\end{equation}
When the rotation operator acts on the vacuum state, it produces only an overall phase. Its contribution can therefore be omitted~\cite{Li:2023ekd}.

Define the complex pair amplitude $\lambda_{A,k}=-e^{2i\varphi_{A,k}}\tanh r_{A,k}$.  Substitution of Eq.~\eqref{eq:compact_squeezed_state} into the Schr\"odinger equation gives the Riccati equation
\begin{equation}
 i\lambda_{A,k}'
 =\Lambda_{A,k}+2\Omega_{A,k}\lambda_{A,k}
 +\Lambda_{A,k}^{*}\lambda_{A,k}^2.
 \label{eq:compact_Riccati}
\end{equation}
This equation is regular at the exact vacuum point $\lambda_{A,k}=0$ and provides an equivalent nonsingular representation of the squeezing dynamics. The real squeezing variables are recovered through $r_{A,k}=\operatorname{arctanh}|\lambda_{A,k}|$ and $\varphi_{A,k}=\arg(-\lambda_{A,k})/2$.

It is useful to introduce the real  pair-creation coefficient $\mathcal A_{A,k}=k(c_A^2-1)/2$ and the rotation frequency $\Omega_{A,k}=k(1+c_A^2)/2$. Using the definition of $\lambda_{A,k}$, Eq.~\eqref{eq:compact_Riccati} then gives the two coupled squeezing equations
\begin{equation}
     r_{A,k}'=\mathcal A_{A,k}\sin(2\varphi_{A,k})
 -h_A\cos(2\varphi_{A,k}),
 \label{eq:compact_r_unified}
\end{equation}
\begin{equation}
    \begin{split}
         \varphi_{A,k}'&=-\Omega_{A,k}
 +\left[\mathcal A_{A,k}\cos(2\varphi_{A,k})
 +h_A\sin(2\varphi_{A,k})\right]\\&\times\coth(2r_{A,k}).
 \label{eq:compact_phi_unified}
    \end{split}
\end{equation}
For scalar perturbations, $(c_A,h_A,r_{A,k},\varphi_{A,k})=(c_s,z_s'/z_s,r_{s,k},\varphi_{s,k})$. For each tensor polarization, they are $(c_T,z_T'/z_T,r_{T,k},\varphi_{T,k})$. These substitutions contain the complete scalar and tensor systems.

We set $M_{\rm Pl}=H_*=1$, $\epsilon_*=0.002$, and $k=0.01$, and use $\delta_{{\rm GB},*}=0$, $0.0025$, and $0.0050$. The first value is the GR reference. All backgrounds satisfy $Q_s>0$, $q_T>0$, $c_s^2>0$, and $c_T^2>0$. The variable $x=-\log(-k\eta)$ runs from the subhorizon regime ($x<0$) through horizon crossing ($x\simeq0$) to the superhorizon regime ($x>0$). For scalar modes the more precise crossing condition is $c_sk\simeq aH$. We use $x=-\log(-k\eta)$ as the evolution variable. Negative $x$ corresponds to the subhorizon regime, $x\simeq0$ approximately marks horizon crossing, and positive $x$ describes superhorizon evolution. For scalar perturbations, the more precise crossing condition is $c_s k\simeq aH$.

Eq.~\eqref{eq:compact_r_unified} separates the two contributions to the squeezing evolution. The real coefficient $\mathcal A_{A,k}=k(c_A^2-1)/2$ describes the contribution associated with a non-unity propagation speed, while $h_A=z_A'/z_A$ describes parametric amplification by the evolving pump field. Both terms contribute in the scalar sector. On the luminal tensor branch, $c_T^2=1$ gives $\mathcal A_{T,k}=0$, so the GB correction enters the tensor squeezing dynamics entirely through $h_T=z_T'/z_T$.

\begin{figure}[h!]
 \centering
  \includegraphics[width=0.85\linewidth]{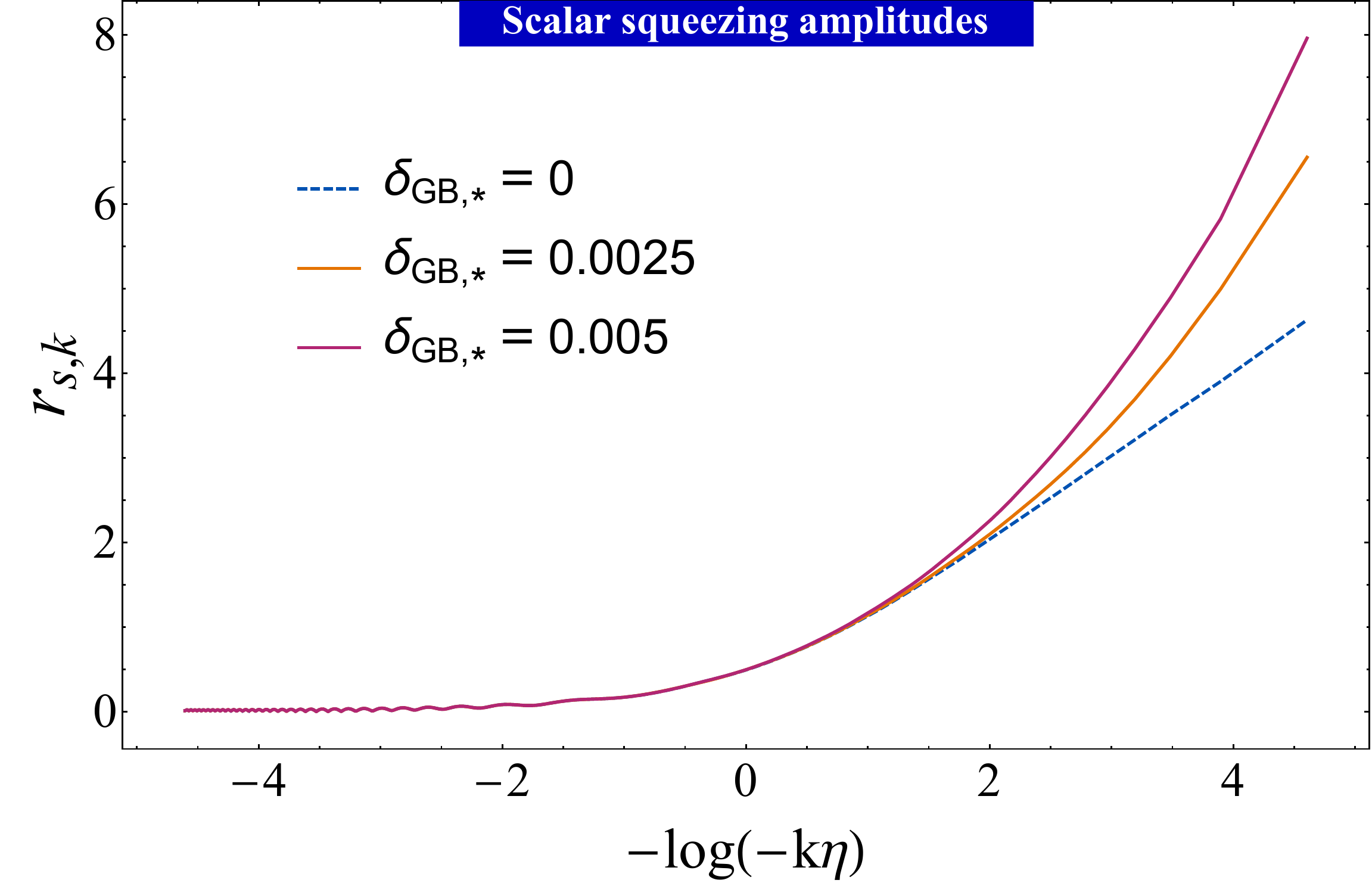}
  \includegraphics[width=0.85\linewidth]{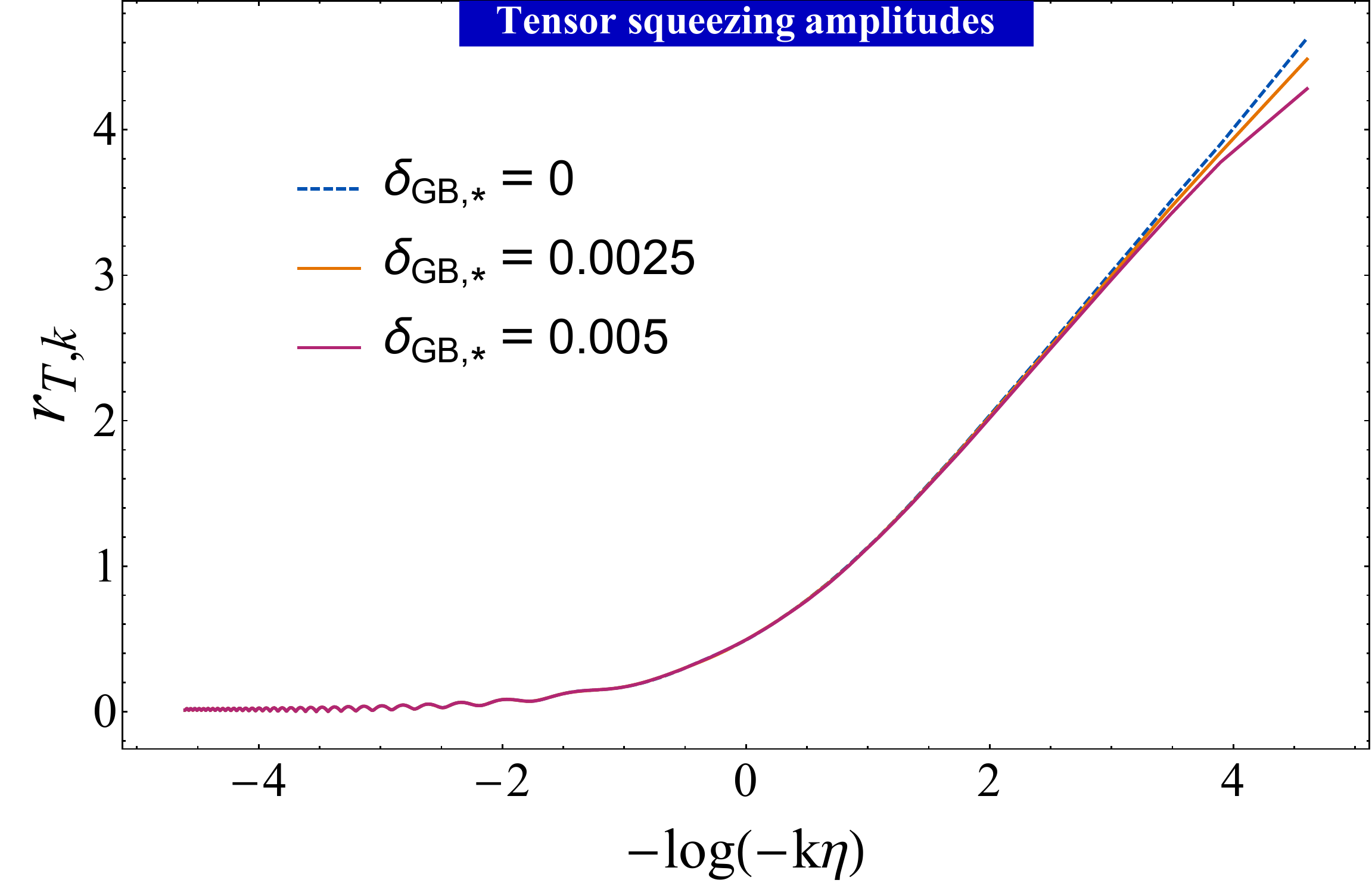}
 \caption{Evolution of the scalar squeezing amplitude $r_{s,k}$ (upper panel) and tensor squeezing amplitude $r_{T,k}$ (lower panel) for different values of $\delta_{{\rm GB},*}$.}
 \label{fig:squeezing_amplitudes}
\end{figure}

Fig.~\ref{fig:squeezing_amplitudes} shows the squeezing amplitudes of the scalar and tensor perturbations. In the upper panel, increasing \(\delta_{{\rm GB},*}\) changes both \(c_s\) and \(h_s=z_s'/z_s\). The scalar curves therefore separate gradually after horizon exit, when the background pump effect dominates over the rapid subhorizon oscillations. Within the parameter range considered here, the scalar squeezing amplitude increases with \(\delta_{{\rm GB},*}\). In the lower panel, the luminal condition gives \(c_T^2=1\), and hence \(\mathcal A_{T,k}=0\). The separation between the tensor curves arises only from \(h_T=z_T'/z_T\), where \(q_T=M_{\rm Pl}^2(1-\delta_{\rm GB})\). The weaker or opposite ordering of the tensor curves is therefore caused by the different pump field rather than by a correction to the tensor propagation speed. The two panels show that luminal tensor propagation does not eliminate the GB correction to tensor squeezing. The squeezing amplitude determines the mean occupation number of correlated mode pairs through \(n_{A,k}=\sinh^2r_{A,k}\). The ordering in Fig.~\ref{fig:squeezing_amplitudes} therefore shows that a positive GB coupling enhances scalar pair production but slightly suppresses tensor pair production for the backgrounds considered here. This opposite response originates from the different evolution of the scalar and tensor pump fields.

The Eq.~\eqref{eq:compact_phi_unified} of squeezed angle, contains a large common rotation. Deep inside the horizon, $c_A^2\simeq1$ and $|h_A|\ll k$, so $\mathcal A_{A,k}\simeq0$, $\Omega_{A,k}\simeq k$, and $\varphi_{A,k}'\simeq-k$. After horizon exit, $r_{A,k}$ grows and $\coth(2r_{A,k})\to1$. The pump term then drives the angle toward the fixed orientations determined approximately by $\sin(2\varphi_{A,k})\simeq0$.

\begin{figure}[h!]
  \centering
  \includegraphics[width=0.85\linewidth]{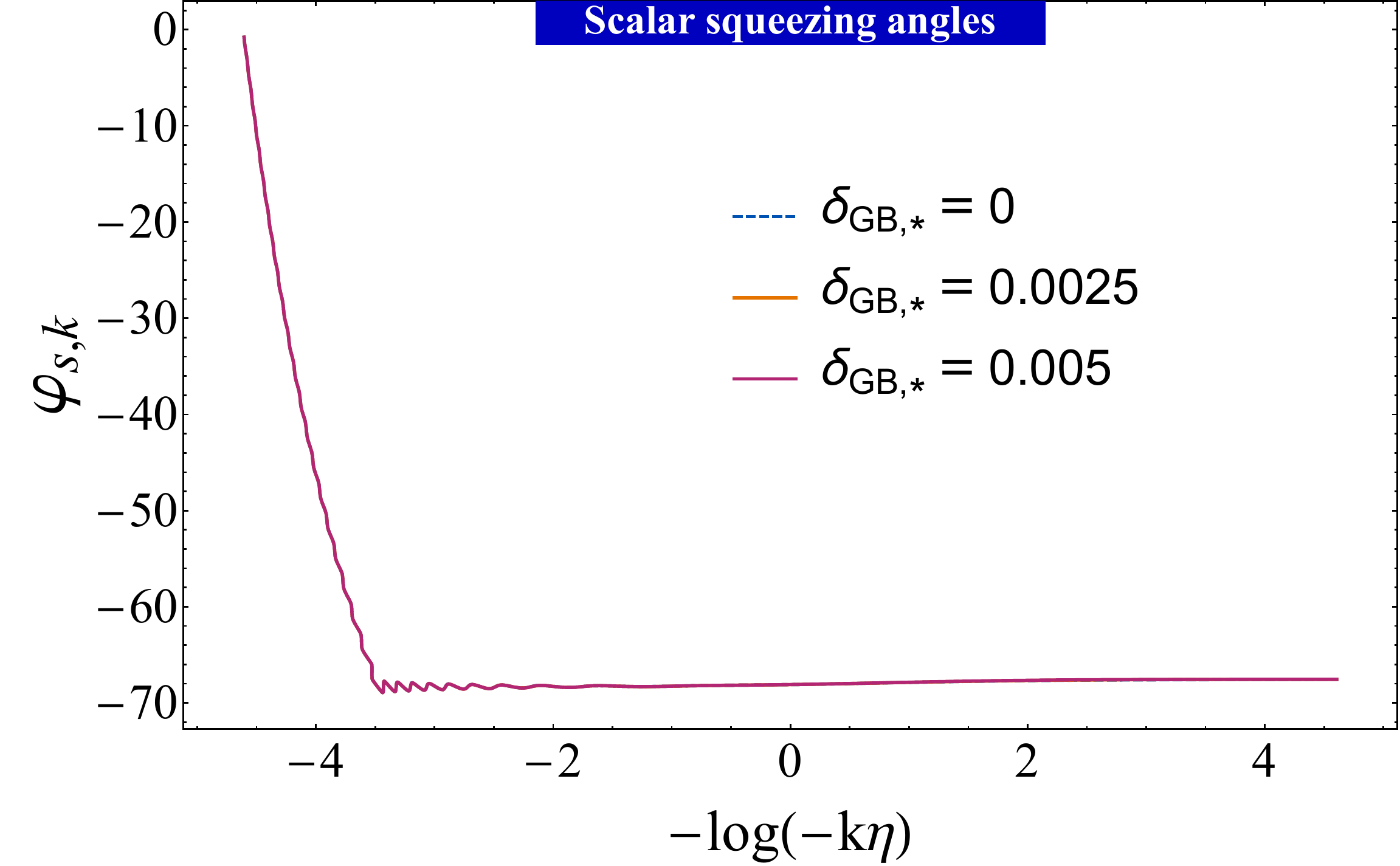}
  \includegraphics[width=0.85\linewidth]{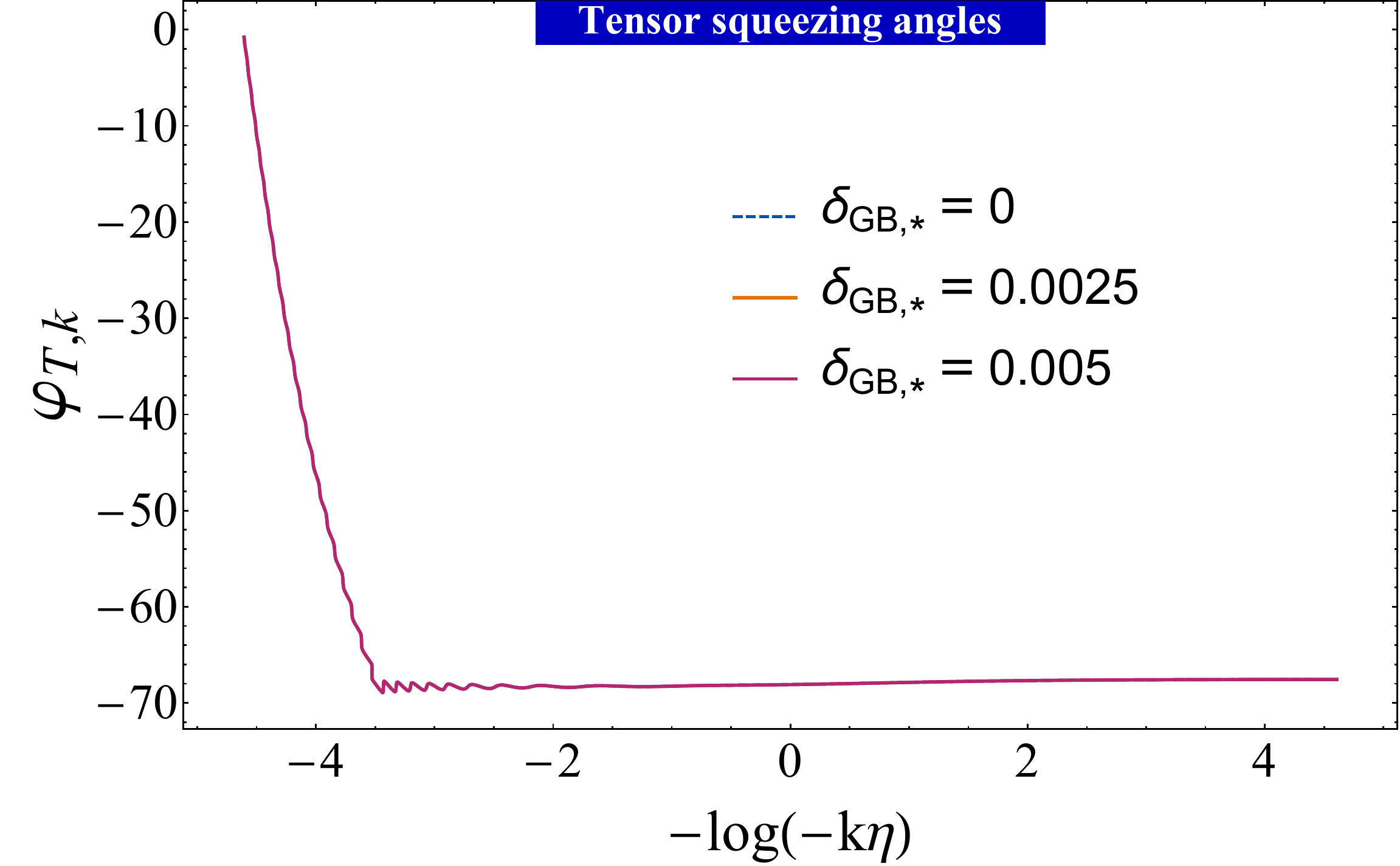}
 \caption{Evolution of the unwrapped scalar squeezing angle $\varphi_{s,k}$ (upper panel) and tensor squeezing angle $\varphi_{T,k}$ (lower panel) for different values of $\delta_{{\rm GB},*}$.}
 \label{fig:squeezing_angles}
\end{figure}

Fig.~\ref{fig:squeezing_angles} shows the unwrapped squeezing angles. The nearly coincident curves in the upper panel arise because the scalar squeezing angle is dominated by the common rotation $-\Omega_{s,k}$. The direct correction from $\mathcal A_{s,k}$ remains small because the stable benchmark solutions keep $c_s^2$ close to unity. In the lower panel, $c_T^2=1$ makes $\mathcal A_{T,k}$ vanish exactly, so the tensor angle receives no direct speed-dependent pair-creation term. The residual GB dependence enters through $h_T$, but it mainly changes the rate at which the solution approaches the superhorizon orientation. It has little effect on that final orientation. The overlap of the curves therefore reflects a common subhorizon phase rotation followed by superhorizon phase locking. It does not imply that the GB correction is absent, as the amplitude evolution in Fig.~\ref{fig:squeezing_amplitudes} remains clearly coupling dependent.

The dependence of the quantum state on \(e^{2i\varphi_{A,k}}\) makes the squeezing angle periodic, such that \(\varphi_{A,k}\sim\varphi_{A,k}+n\pi\). The small coupling-dependent phase shift can be extracted through the following principal-value angular residual
\begin{equation}
 \Delta\varphi_{A,k}=\frac12\operatorname{Arg}\!\left\{
 \exp\!\left[2i\left(\varphi_{A,k}-\varphi_{A,k}^{\rm GR}\right)\right]\right\}.
 \label{eq:compact_phase_residual}
\end{equation}

\begin{figure}[h!]
  \centering
  \includegraphics[width=0.85\linewidth]{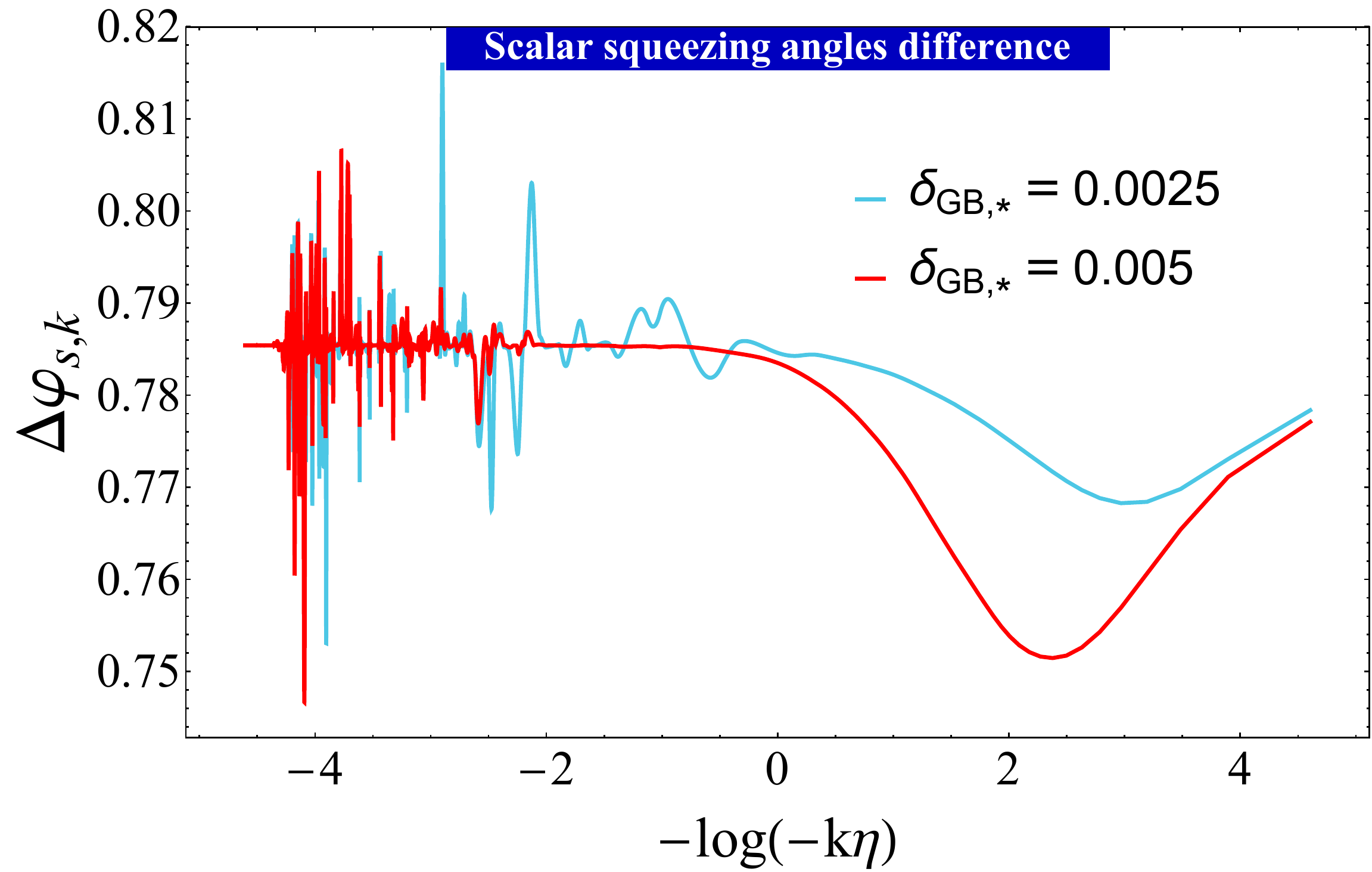}
  \includegraphics[width=0.85\linewidth]{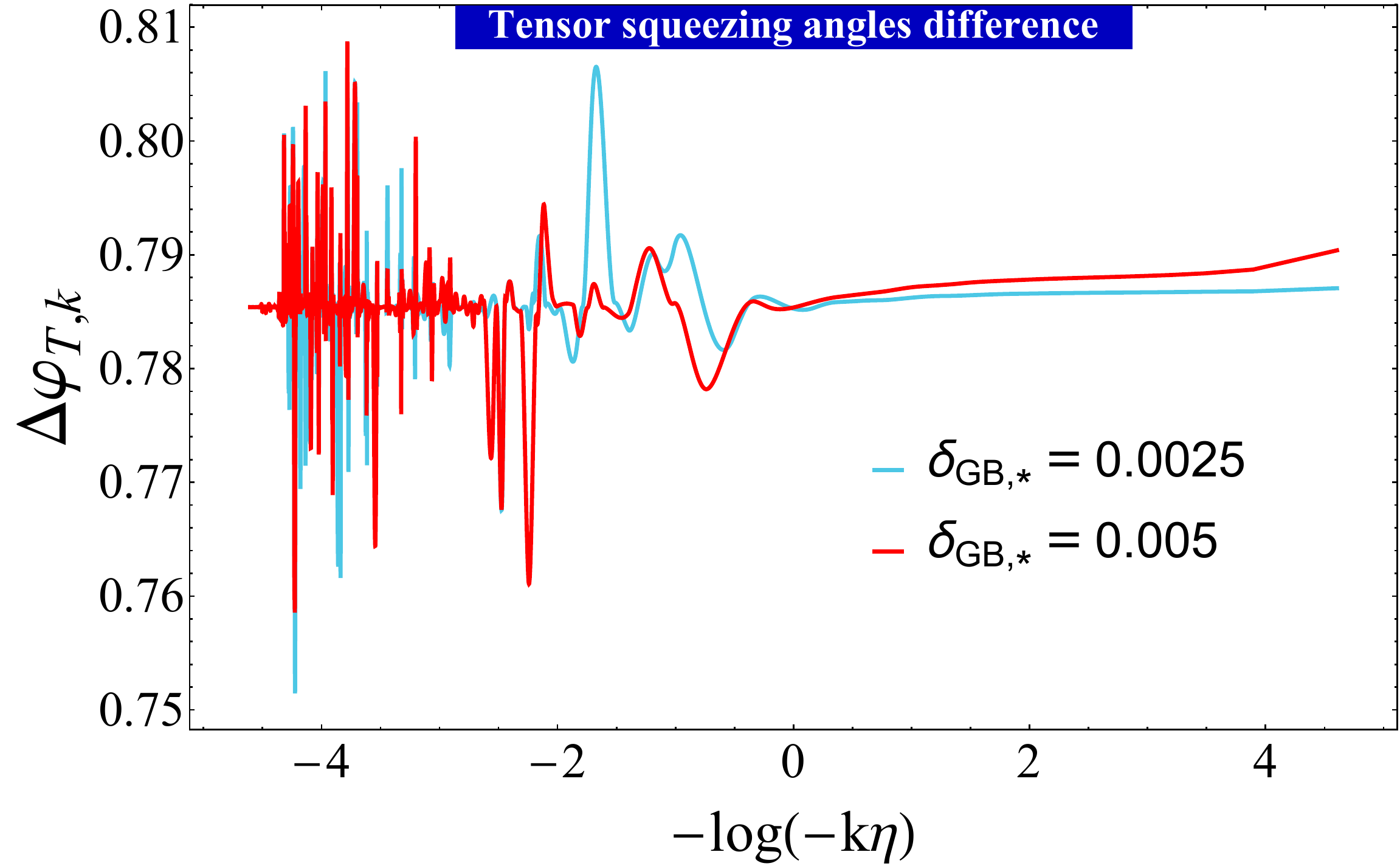}
 \caption{Principal squeezing angle residual relative to general relativity in the scalar sector, $\Delta\varphi_{s,k}$ (upper panel), and tensor sector, $\Delta\varphi_{T,k}$ (lower panel).}
 \label{fig:angle_residuals}
\end{figure}

Fig.~\ref{fig:angle_residuals} displays the coupling-dependent phase information hidden by the large common rotation in Fig.~\ref{fig:squeezing_angles}. The scalar residual receives contributions from both the sound-speed correction and the modified scalar pump field, whereas the tensor residual is generated only by $h_T$ because $c_T^2=1$. In both sectors, the residuals remain finite but are much smaller than the accumulated unwrapped rotation. The GB coupling therefore produces a genuine phase correction without qualitatively changing the phase-locking behavior. The dominant state-space separation is carried by the squeezing amplitude, while $\Delta\varphi_{A,k}$ provides a smaller phase-sensitive contribution that must be retained in the circuit complexity.

The present construction is particularly suited to circuit complexity. The reference state and the target two-mode squeezed state are already specified, and the latter is completely characterized by $r_{A,k}$ and $\varphi_{A,k}$. 

\section{Circuit complexity of the squeezed perturbations}
\label{sec:circuit_complexity}

Circuit complexity measures the shortest allowed circuit that prepares a target state from a chosen reference state. Its geometric formulation represents the circuit by a path on the manifold of unitary transformations and minimizes a prescribed cost along that path~\cite{Nielsen:2005mkt,Dowling:2006tnk}. This construction has been extended to Gaussian states and quantum field theory using wave functions, covariance matrices, and continuous gate sets~\cite{Jefferson:2017sdb,Chapman:2017rqy,Chapman:2018hou,Hackl:2018ptj}. Related studies have examined the role of the reference state, penalty factors, gate scale, and alternative Gaussian state geometries~\cite{Bhattacharyya:2018bbv,Alves:2018qfv,Camargo:2018eof,Ali:2018fcz}. Circuit complexity has also been applied to squeezed cosmological perturbations, where the Bunch--Davies vacuum provides a natural reference and the evolved two-mode squeezed state provides the target~\cite{Bhattacharyya:2020art,Li:2023ekd}.

For each sector $A=s,T$, we take the un-squeezed vacuum at the initial time as the reference state and the state in Eq.~\eqref{eq:compact_squeezed_state} as the target. We adopt the unpenalized Nielsen $F_2$ cost evaluated from the Gaussian wave function. In the conventions of the present squeeze operator, the circuit complexity is~\cite{Bhattacharyya:2020art}
\begin{equation}
 \begin{split}
 \mathcal C_{2,A}(k)&=\frac{1}{\sqrt{2}}\Bigg\{
 \left[\ln\left|\frac{1+e^{-2i\varphi_{A,k}}\tanh r_{A,k}}
 {1-e^{-2i\varphi_{A,k}}\tanh r_{A,k}}\right|\right]^2\\
 &+\left[\tan^{-1}\!\left(2\sin(2\varphi_{A,k})
 \sinh r_{A,k}\cosh r_{A,k}\right)\right]^2
 \Bigg\}^{1/2}.
 \label{eq:circuit_complexity_F2}
 \end{split}
\end{equation}
This expression keeps both pieces of information carried by the squeezed state. The logarithmic term is mainly controlled by the squeezing amplitude, while the bounded angular term records the orientation of the squeezed ellipse in phase space. In the superhorizon phase-locked regime, where $\sin(2\varphi_{A,k})\simeq0$, Eq.~\eqref{eq:circuit_complexity_F2} reduces approximately to $\mathcal C_{2,A}(k)\simeq\sqrt{2}\,r_{A,k}$. The late-time ordering of the circuit complexity curves is therefore expected to follow the ordering of the squeezing amplitudes. The finite angular residual changes the detailed curve but does not reverse this ordering in the present parameter range. Eq. \eqref{eq:circuit_complexity_F2} depends on specific choices, including the gate set, cost function, reference frequency, and covariance-matrix prescription. These choices can change its numerical normalization and may remove the explicit dependence on the squeezing angle. Eq.~\eqref{eq:circuit_complexity_F2} is therefore used consistently for both scalar and tensor sectors.

The numerical curves are obtained by inserting the solutions $(r_{s,k},\varphi_{s,k})$ and $(r_{T,k},\varphi_{T,k})$ from Eqs.~\eqref{eq:compact_r_unified} and \eqref{eq:compact_phi_unified} into Eq.~\eqref{eq:circuit_complexity_F2}. The same reference state, cost function, comoving mode, and GB benchmarks are used in both sectors. This common prescription makes the scalar--tensor comparison sensitive only to their different propagation and pump coefficients. Since the squeezing angles in Fig.~\ref{fig:squeezing_angles} remain nearly phase locked, the main difference in circuit complexity is expected to be determined by the squeezing amplitudes, while the squeezing angle contribution remains small and bounded.

\begin{figure}[t]
    \centering
    \includegraphics[width=0.85\linewidth]{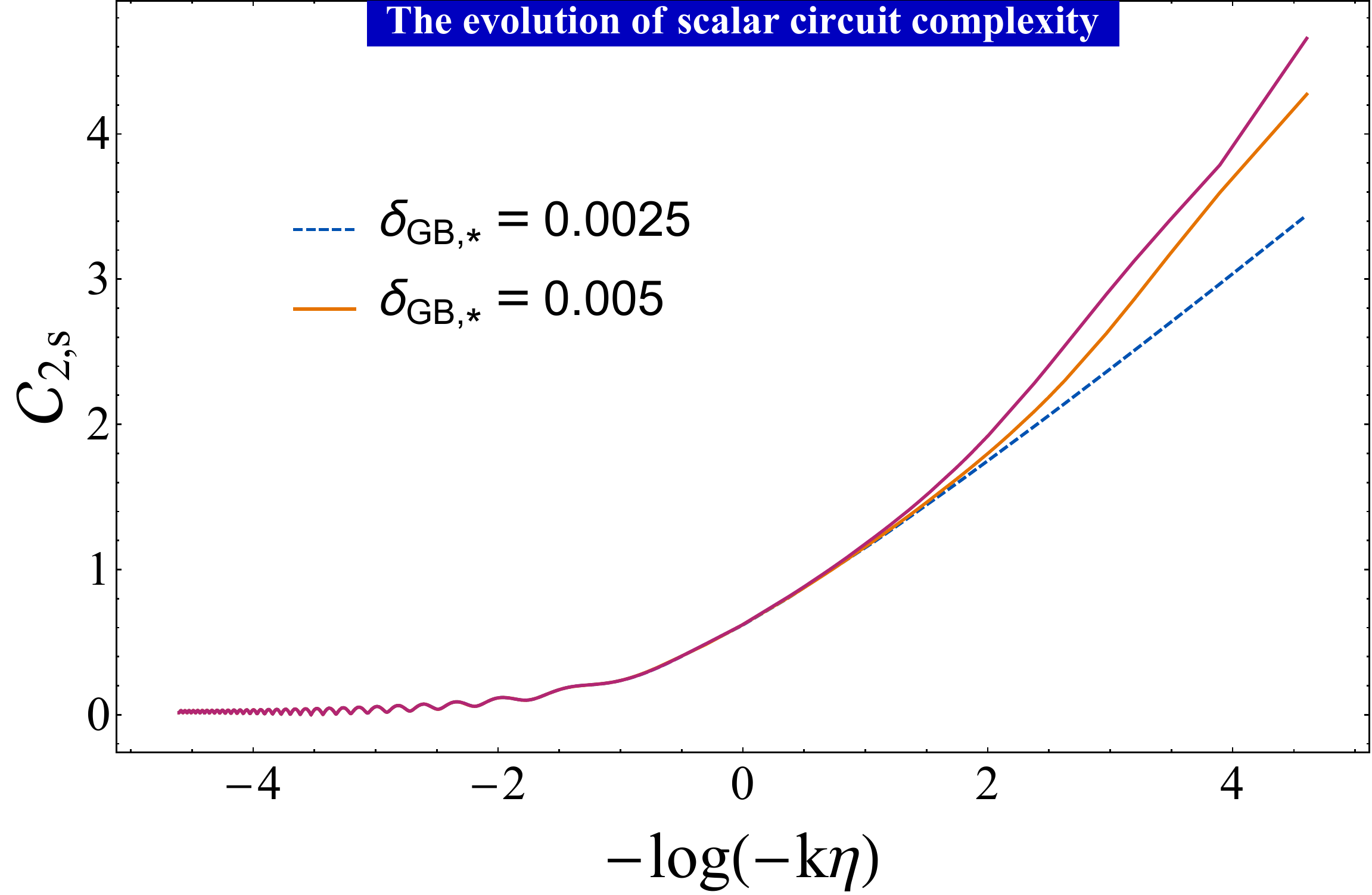}
    \includegraphics[width=0.85\linewidth]{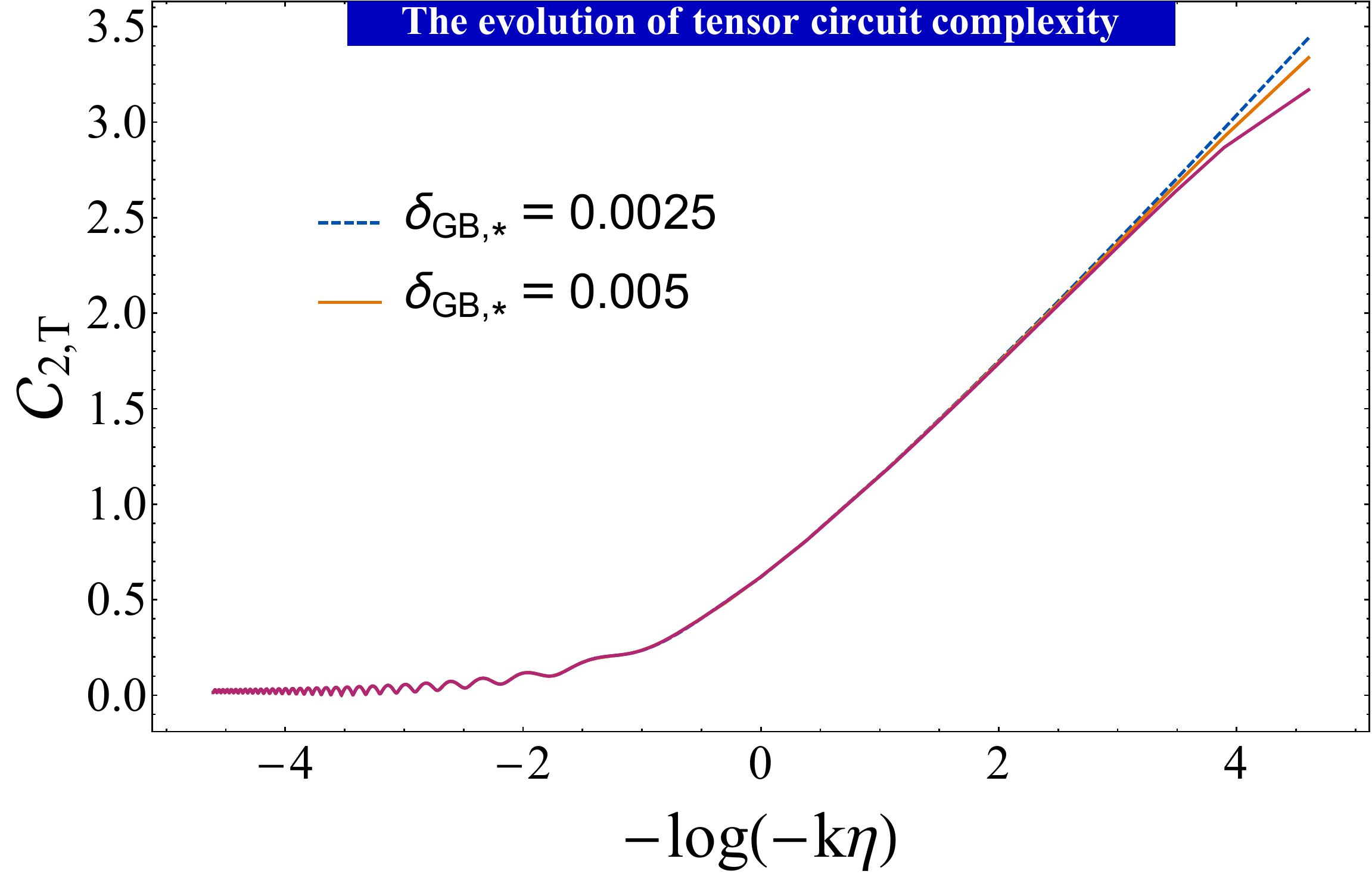}
    \caption{Evolution of the Nielsen $F_2$ circuit complexity for scalar perturbations and tensor perturbations. The horizontal axis $-\log(-k\eta)$ runs from the subhorizon regime on the left to the superhorizon regime on the right. The dashed blue curves denote the GR limit $\delta_{{\rm GB},*}=0$, while the orange and magenta curves correspond to $\delta_{{\rm GB},*}=0.0025$ and $0.005$, respectively. The scalar complexity is enhanced by increasing positive GB coupling, whereas the tensor complexity is slightly suppressed.}
    \label{fig:circuit_complexities}
\end{figure}

Fig.~\ref{fig:circuit_complexities} presents the numerical results for the Nielsen $F_2$ circuit complexity. In the upper panel, the scalar circuit complexity remains close to zero in the subhorizon regime because the state is close to the unsqueezed vacuum reference state. Small oscillations appear when the pump field begins to compete with the mode frequency, and the subsequent growth of $r_{s,k}$ after horizon exit drives a rapid increase in the circuit complexity. For the positive benchmark couplings considered here, the late-time ordering $\mathcal C_{2,s}(0.005)>\mathcal C_{2,s}(0.0025)>\mathcal C_{2,s}(0)$ follows the ordering of the scalar squeezing amplitude in Fig.~\ref{fig:squeezing_amplitudes}. The lower panel shows a smaller separation in the tensor sector. On the luminal branch, $c_T^2=1$ gives $\mathcal A_{T,k}=0$, but the time-dependent tensor kinetic coefficient $q_T$ keeps the pump field $h_T=z_T'/z_T$ nontrivial. The resulting pump-field correction slightly reduces $r_{T,k}$ and gives $\mathcal C_{2,T}(0.005)<\mathcal C_{2,T}(0.0025)<\mathcal C_{2,T}(0)$ at late times. The opposite scalar and tensor orderings therefore arise from the different background dependence of $z_s$ and $z_T$, rather than from a direct correction to the tensor propagation speed. The squeezing-angle contribution changes the detailed shapes of the curves but remains subleading because the squeezing angles are nearly phase locked. Relative to their corresponding vacuum reference states, the same positive GB coupling increases the preparation cost of the scalar Gaussian state but slightly decreases that of the tensor Gaussian state, making the scalar sector more sensitive to the EGB background and the luminal tensor sector a cleaner probe of the pump-field mechanism.

\section{Conclusions and outlook}
\label{sec:conclusion}

We studied the scalar and tensor two-mode squeezed states in the luminal branch of EGB inflation and evaluated their Nielsen $F_2$ circuit complexities. The quadratic actions lead to different kinetic coefficients, propagation speeds, and pump fields in the two sectors. Imposing $c_T^2=1$ removes the direct tensor-speed correction. However, the time-dependent coefficient $q_T$ continues to modify $z_T'/z_T$. Luminal tensor propagation therefore does not imply general-relativistic evolution of the tensor quantum state.

The GB coupling affects the squeezing amplitudes more strongly than the squeezing angles. Deep inside the horizon, both sectors are dominated by a large common phase rotation. After horizon exit, the pump fields amplify the states and drive their squeezing angles toward nearly phase-locked orientations. The principal angular residuals remain finite but small. The resulting circuit complexity is therefore controlled mainly by the squeezing amplitude, with a smaller phase-sensitive contribution from the squeezing angle.

For the positive benchmark couplings considered here, increasing $\delta_{{\rm GB},*}$ enhances the scalar squeezing amplitude and circuit complexity but slightly suppresses their tensor counterparts. The opposite ordering originates from the different background dependence of $Q_s$ and $q_T$ and their associated pump fields. Relative to their corresponding vacuum reference states, the same GB coupling therefore increases the preparation cost of the scalar Gaussian state but decreases that of the tensor Gaussian state.

Within the circuit prescription adopted here, the larger scalar preparation cost indicates that the scalar Gaussian state is more sensitive to the GB coupling. The scalar sector is therefore better suited to reflecting changes in the quantum-state geometry. By contrast, the smaller tensor correction provides a cleaner probe of the pump-field mechanism because the luminal condition removes the direct tensor-speed contribution. Neither sector is universally preferable. Their opposite responses make the combined scalar--tensor comparison more informative than either sector alone. The circuit complexity considered here is a state-space diagnostic rather than a direct cosmological observable. Its numerical value depends on the background solution, reference state, gate set, and cost function. Future work may test whether the scalar--tensor complexity splitting persists in other stable EGB backgrounds and under alternative circuit prescriptions. It may also explore its possible relation to conventional inflationary correlation functions.

\subsection*{Acknowledgments}
This work is supported by the National Natural Science Foundation of China under Grant No.12665017, the Guizhou Provincial Natural Science Foundation (Grant No. 2025YJSKYJJ178), the Natural Science Foundation of Guizhou Minzu University (Grant No. GZMUBS7), the National Natural Science Foundation of China (Grant Nos. 12665009 and 12265010), the Hunan Provincial Department of Education Project (Grant No. 25B0480), and the Guizhou Provincial Department of Science and Technology (Grant Nos. MS[2025]219 and CXTD[2025]030).

\bibliography{refs.bib}

@article{Guo:2009uk,
    author = "Guo, Zong-Kuan and Schwarz, Dominik J.",
    title = "{Power spectra from an inflaton coupled to the Gauss-Bonnet term}",
    eprint = "0907.0427",
    archivePrefix = "arXiv",
    primaryClass = "hep-th",
    reportNumber = "BI-TP-2009-15",
    doi = "10.1103/PhysRevD.80.063523",
    journal = "Phys. Rev. D",
    volume = "80",
    pages = "063523",
    year = "2009"
}

@article{Guo:2010jr,
    author = "Guo, Zong-Kuan and Schwarz, Dominik J.",
    title = "{Slow-roll inflation with a Gauss-Bonnet correction}",
    eprint = "1001.1897",
    archivePrefix = "arXiv",
    primaryClass = "hep-th",
    reportNumber = "BI-TP-2010-01",
    doi = "10.1103/PhysRevD.81.123520",
    journal = "Phys. Rev. D",
    volume = "81",
    pages = "123520",
    year = "2010"
}

@article{Kobayashi:2011nu,
    author = "Kobayashi, Tsutomu and Yamaguchi, Masahide and Yokoyama, Jun'ichi",
    title = "{Generalized G-inflation: Inflation with the most general second-order field equations}",
    eprint = "1105.5723",
    archivePrefix = "arXiv",
    primaryClass = "hep-th",
    reportNumber = "KUNS-2339, RESCEU-9-11",
    doi = "10.1143/PTP.126.511",
    journal = "Prog. Theor. Phys.",
    volume = "126",
    pages = "511--529",
    year = "2011"
}

@article{Li:2023ekd,
    author = "Li, Tao and Liu, Lei-Hua",
    title = "{Cosmological complexity of the modified dispersion relation}",
    eprint = "2309.01595",
    archivePrefix = "arXiv",
    primaryClass = "gr-qc",
    doi = "10.1016/j.physletb.2024.138728",
    journal = "Phys. Lett. B",
    volume = "854",
    pages = "138728",
    year = "2024"
}

@article{Jiang:2013gza,
    author = "Jiang, Peng-Xu and Hu, Jian-Wei and Guo, Zong-Kuan",
    title = "{Inflation coupled to a Gauss-Bonnet term}",
    eprint = "1310.5579",
    archivePrefix = "arXiv",
    primaryClass = "hep-th",
    doi = "10.1103/PhysRevD.88.123508",
    journal = "Phys. Rev. D",
    volume = "88",
    pages = "123508",
    year = "2013"
}

@article{Hwang:2005hb,
    author = "Hwang, Jai-chan and Noh, Hyerim",
    title = "{Classical evolution and quantum generation in generalized gravity theories including string corrections and tachyon: Unified analyses}",
    eprint = "gr-qc/0412126",
    archivePrefix = "arXiv",
    doi = "10.1103/PhysRevD.71.063536",
    journal = "Phys. Rev. D",
    volume = "71",
    pages = "063536",
    year = "2005"
}

@article{DeFelice:2011uc,
    author = "De Felice, Antonio and Tsujikawa, Shinji",
    title = "{Inflationary non-Gaussianities in the most general second-order scalar-tensor theories}",
    eprint = "1107.3917",
    archivePrefix = "arXiv",
    primaryClass = "gr-qc",
    doi = "10.1103/PhysRevD.84.083504",
    journal = "Phys. Rev. D",
    volume = "84",
    pages = "083504",
    year = "2011"
}

@article{Nielsen:2005mkt,
    author = "Nielsen, Michael A.",
    title = "{A geometric approach to quantum circuit lower bounds}",
    eprint = "quant-ph/0502070",
    archivePrefix = "arXiv",
    doi = "10.26421/QIC6.3-2",
    journal = "Quant. Inf. Comput.",
    volume = "6",
    number = "3",
    pages = "213--262",
    year = "2006"
}

@article{Dowling:2006tnk,
    author = "Dowling, Mark R. and Nielsen, Michael A.",
    title = "{The geometry of quantum computation}",
    eprint = "quant-ph/0701004",
    archivePrefix = "arXiv",
    doi = "10.26421/QIC8.10-1",
    journal = "Quant. Inf. Comput.",
    volume = "8",
    number = "10",
    pages = "0861--0899",
    year = "2008"
}

@article{Jefferson:2017sdb,
    author = "Jefferson, Ro and Myers, Robert C.",
    title = "{Circuit complexity in quantum field theory}",
    eprint = "1707.08570",
    archivePrefix = "arXiv",
    primaryClass = "hep-th",
    doi = "10.1007/JHEP10(2017)107",
    journal = "JHEP",
    volume = "10",
    pages = "107",
    year = "2017"
}

@article{Chapman:2017rqy,
    author = "Chapman, Shira and Heller, Michal P. and Marrochio, Hugo and Pastawski, Fernando",
    title = "{Toward a Definition of Complexity for Quantum Field Theory States}",
    eprint = "1707.08582",
    archivePrefix = "arXiv",
    primaryClass = "hep-th",
    doi = "10.1103/PhysRevLett.120.121602",
    journal = "Phys. Rev. Lett.",
    volume = "120",
    number = "12",
    pages = "121602",
    year = "2018"
}

@article{Chapman:2018hou,
    author = "Chapman, Shira and Eisert, Jens and Hackl, Lucas and Heller, Michal P. and Jefferson, Ro and Marrochio, Hugo and Myers, Robert C.",
    title = "{Complexity and entanglement for thermofield double states}",
    eprint = "1810.05151",
    archivePrefix = "arXiv",
    primaryClass = "hep-th",
    doi = "10.21468/SciPostPhys.6.3.034",
    journal = "SciPost Phys.",
    volume = "6",
    number = "3",
    pages = "034",
    year = "2019"
}

@article{Hackl:2018ptj,
    author = "Hackl, Lucas and Myers, Robert C.",
    title = "{Circuit complexity for free fermions}",
    eprint = "1803.10638",
    archivePrefix = "arXiv",
    primaryClass = "hep-th",
    reportNumber = "IGC-18/3-1, IGC-18-3-1",
    doi = "10.1007/JHEP07(2018)139",
    journal = "JHEP",
    volume = "07",
    pages = "139",
    year = "2018"
}

@article{Bhattacharyya:2018bbv,
    author = "Bhattacharyya, Arpan and Shekar, Arvind and Sinha, Aninda",
    title = "{Circuit complexity in interacting QFTs and RG flows}",
    eprint = "1808.03105",
    archivePrefix = "arXiv",
    primaryClass = "hep-th",
    reportNumber = "YITP-18-89",
    doi = "10.1007/JHEP10(2018)140",
    journal = "JHEP",
    volume = "10",
    pages = "140",
    year = "2018"
}

@article{Alves:2018qfv,
    author = "Alves, Daniel W. F. and Camilo, Giancarlo",
    title = "{Evolution of complexity following a quantum quench in free field theory}",
    eprint = "1804.00107",
    archivePrefix = "arXiv",
    primaryClass = "hep-th",
    doi = "10.1007/JHEP06(2018)029",
    journal = "JHEP",
    volume = "06",
    pages = "029",
    year = "2018"
}

@article{Camargo:2018eof,
    author = "Camargo, Hugo A. and Caputa, Pawel and Das, Diptarka and Heller, Michal P. and Jefferson, Ro",
    title = "{Complexity as a novel probe of quantum quenches: universal scalings and purifications}",
    eprint = "1807.07075",
    archivePrefix = "arXiv",
    primaryClass = "hep-th",
    doi = "10.1103/PhysRevLett.122.081601",
    journal = "Phys. Rev. Lett.",
    volume = "122",
    number = "8",
    pages = "081601",
    year = "2019"
}

@article{Ali:2018fcz,
    author = "Ali, Tibra and Bhattacharyya, Arpan and Shajidul Haque, S. and Kim, Eugene H. and Moynihan, Nathan",
    title = "{Time Evolution of Complexity: A Critique of Three Methods}",
    eprint = "1810.02734",
    archivePrefix = "arXiv",
    primaryClass = "hep-th",
    reportNumber = "YITP-18-111",
    doi = "10.1007/JHEP04(2019)087",
    journal = "JHEP",
    volume = "04",
    pages = "087",
    year = "2019"
}

@article{Bhattacharyya:2020art,
    author = "Bhattacharyya, Arpan and Chemissany, Wissam and Haque, S. Shajidul and Murugan, Jeff and Yan, Bin",
    title = "{The Multi-faceted Inverted Harmonic Oscillator: Chaos and Complexity}",
    eprint = "2007.01232",
    archivePrefix = "arXiv",
    primaryClass = "hep-th",
    reportNumber = "LA-UR-20-24798",
    doi = "10.21468/SciPostPhysCore.4.1.002",
    journal = "SciPost Phys. Core",
    volume = "4",
    pages = "002",
    year = "2021"
}

@article{Grishchuk:1990bj,
    author = "Grishchuk, L. P. and Sidorov, Yu. V.",
    title = "{Squeezed quantum states of relic gravitons and primordial density fluctuations}",
    doi = "10.1103/PhysRevD.42.3413",
    journal = "Phys. Rev. D",
    volume = "42",
    pages = "3413--3421",
    year = "1990"
}

@article{Albrecht:1992kf,
    author = "Albrecht, Andreas and Ferreira, Pedro and Joyce, Michael and Prokopec, Tomislav",
    title = "{Inflation and squeezed quantum states}",
    eprint = "astro-ph/9303001",
    archivePrefix = "arXiv",
    reportNumber = "IMPERIAL-TP-92-93-21",
    doi = "10.1103/PhysRevD.50.4807",
    journal = "Phys. Rev. D",
    volume = "50",
    pages = "4807--4820",
    year = "1994"
}

@article{Liu:2021nzx,
    author = "Liu, Lei-Hua and Li, Ai-Chen",
    title = "{Complexity of non-trivial sound speed in inflation}",
    eprint = "2102.12014",
    archivePrefix = "arXiv",
    primaryClass = "gr-qc",
    doi = "10.1016/j.dark.2022.101123",
    journal = "Phys. Dark Univ.",
    volume = "37",
    pages = "101123",
    year = "2022"
}

@article{Li:2024kfm,
    author = "Li, Tao and Liu, Lei-Hua",
    title = "{Inflationary Krylov complexity}",
    eprint = "2401.09307",
    archivePrefix = "arXiv",
    primaryClass = "hep-th",
    doi = "10.1007/JHEP04(2024)123",
    journal = "JHEP",
    volume = "04",
    pages = "123",
    year = "2024"
}

@article{Li:2026jxx,
    author = "Li, Tao and Fu, Hai-Bing",
    title = "{Circuit and Krylov complexity of primordial perturbations of modified gravity in inflation}",
    eprint = "2607.09408",
    archivePrefix = "arXiv",
    primaryClass = "hep-th",
    journal = "",
    month = "7",
    year = "2026"
}

@article{Kawai:1999pw,
    author = "Kawai, Shinsuke and Soda, Jiro",
    title = "{Evolution of fluctuations during graceful exit in string cosmology}",
    eprint = "gr-qc/9903017",
    archivePrefix = "arXiv",
    doi = "10.1016/S0370-2693(99)00736-4",
    journal = "Phys. Lett. B",
    volume = "460",
    pages = "41--46",
    year = "1999"
}

@article{Satoh:2008ck,
    author = "Satoh, Masaki and Soda, Jiro",
    title = "{Higher Curvature Corrections to Primordial Fluctuations in Slow-roll Inflation}",
    eprint = "0806.4594",
    archivePrefix = "arXiv",
    primaryClass = "astro-ph",
    doi = "10.1088/1475-7516/2008/09/019",
    journal = "JCAP",
    volume = "09",
    pages = "019",
    year = "2008"
}

@article{Kanti:2015pda,
    author = "Kanti, Panagiota and Gannouji, Radouane and Dadhich, Naresh",
    title = "{Gauss-Bonnet Inflation}",
    eprint = "1503.01579",
    archivePrefix = "arXiv",
    primaryClass = "hep-th",
    doi = "10.1103/PhysRevD.92.041302",
    journal = "Phys. Rev. D",
    volume = "92",
    number = "4",
    pages = "041302",
    year = "2015"
}

@article{Martin:2015qta,
    author = "Martin, Jerome and Vennin, Vincent",
    title = "{Quantum Discord of Cosmic Inflation: Can we Show that CMB Anisotropies are of Quantum-Mechanical Origin?}",
    eprint = "1510.04038",
    archivePrefix = "arXiv",
    primaryClass = "astro-ph.CO",
    doi = "10.1103/PhysRevD.93.023505",
    journal = "Phys. Rev. D",
    volume = "93",
    number = "2",
    pages = "023505",
    year = "2016"
}

@article{Haque:2021hyw,
    author = "Haque, S. Shajidul and Jana, Chandan and Underwood, Bret",
    title = "{Operator complexity for quantum scalar fields and cosmological perturbations}",
    eprint = "2110.08356",
    archivePrefix = "arXiv",
    primaryClass = "hep-th",
    doi = "10.1103/PhysRevD.106.063510",
    journal = "Phys. Rev. D",
    volume = "106",
    number = "6",
    pages = "063510",
    year = "2022"
}

@article{Lehners:2020pem,
    author = "Lehners, Jean-Luc and Quintin, Jerome",
    title = "{Quantum Circuit Complexity of Primordial Perturbations}",
    eprint = "2012.04911",
    archivePrefix = "arXiv",
    primaryClass = "hep-th",
    doi = "10.1103/PhysRevD.103.063527",
    journal = "Phys. Rev. D",
    volume = "103",
    number = "6",
    pages = "063527",
    year = "2021"
}

@article{Yi:2018gse,
    author = "Yi, Zhu and Gong, Yungui and Sabir, Mudassar",
    title = "{Inflation with Gauss-Bonnet coupling}",
    eprint = "1804.09116",
    archivePrefix = "arXiv",
    primaryClass = "gr-qc",
    doi = "10.1103/PhysRevD.98.083521",
    journal = "Phys. Rev. D",
    volume = "98",
    number = "8",
    pages = "083521",
    year = "2018"
}

@article{Koh:2014bka,
    author = "Koh, Seoktae and Lee, Bum-Hoon and Lee, Wonwoo and Tumurtushaa, Gansukh",
    title = "{Observational constraints on slow-roll inflation coupled to a Gauss-Bonnet term}",
    eprint = "1404.6096",
    archivePrefix = "arXiv",
    primaryClass = "gr-qc",
    doi = "10.1103/PhysRevD.90.063527",
    journal = "Phys. Rev. D",
    volume = "90",
    number = "6",
    pages = "063527",
    year = "2014"
}

@article{Mudrunka:2023wxy,
    author = "Mudrunka, Kamil and Nakayama, Kazunori",
    title = "{Probing Gauss-Bonnet-corrected inflation with gravitational waves}",
    eprint = "2312.15766",
    archivePrefix = "arXiv",
    primaryClass = "astro-ph.CO",
    reportNumber = "TU-1217, KEK-QUP-0036",
    doi = "10.1088/1475-7516/2024/05/069",
    journal = "JCAP",
    volume = "05",
    pages = "069",
    year = "2024"
}

@article{Satoh:2010ep,
    author = "Satoh, Masaki",
    title = "{Slow-roll Inflation with the Gauss-Bonnet and Chern-Simons Corrections}",
    eprint = "1008.2724",
    archivePrefix = "arXiv",
    primaryClass = "astro-ph.CO",
    doi = "10.1088/1475-7516/2010/11/024",
    journal = "JCAP",
    volume = "11",
    pages = "024",
    year = "2010"
}

@article{Akbarieh:2021vhv,
    author = "Akbarieh, Amin Rezaei and Kazempour, Sobhan and Shao, Lijing",
    title = "{Cosmological perturbations in Gauss-Bonnet quasi-dilaton massive gravity}",
    eprint = "2105.03744",
    archivePrefix = "arXiv",
    primaryClass = "gr-qc",
    doi = "10.1103/PhysRevD.103.123518",
    journal = "Phys. Rev. D",
    volume = "103",
    pages = "123518",
    year = "2021"
}

@article{Odintsov:2020sqy,
    author = "Odintsov, S. D. and Oikonomou, V. K. and Fronimos, F. P.",
    title = "{Rectifying Einstein-Gauss-Bonnet Inflation in View of GW170817}",
    eprint = "2003.13724",
    archivePrefix = "arXiv",
    primaryClass = "gr-qc",
    doi = "10.1016/j.nuclphysb.2020.115135",
    journal = "Nucl. Phys. B",
    volume = "958",
    pages = "115135",
    year = "2020"
}

@article{Li:2026cgb,
    author = "Li, Tao and Fu, Hai-Bing and Liu, Lei-Hua",
    title = "{Circuit complexity of coupled four-mode squeezed states in two-field inflation}",
    eprint = "2610.38254",
    archivePrefix = "arXiv",
    primaryClass = "hep-th",
     journal = "",
    month = "10",
    year = "2026"
}

@article{Li:2021kfq,
    author = "Li, Ai-chen and Li, Xin-Fei and Zeng, Ding-fang and Liu, Lei-Hua",
    title = "{Cosmological complexity in K-essence}",
    eprint = "2102.12939",
    archivePrefix = "arXiv",
    primaryClass = "gr-qc",
    doi = "10.1016/j.dark.2024.101422",
    journal = "Phys. Dark Univ.",
    volume = "43",
    pages = "101422",
    year = "2024"
}

\end{document}